\documentclass[11pt]{article}

\usepackage{amssymb}

\def\one{{\mathbf 1}}
\def\id{\mbox{\it id}}
\def\Box{\square}

\def\text#1{\mbox{\rm #1\ }}
\def\stackunder#1#2{\mathrel{\mathop{#2}\limits_{#1}}}

\newcommand{\CC}{\mathbb{C}}

\title{Associated quantum vector bundles and symplectic structure on a 
       quantum plane\thanks{Contribution to the Proceedings of the
       $8^{th}$ International Colloquium {\bf Quantum Groups and Integrable
       Systems}, Prague, June 17-19, 1999.}
    \vspace{0.5cm}
}

\author{R. Coquereaux${}^1$\thanks{~Email: coque@cpt.univ-mrs.fr}$\;$,
        A. O. Garc\'{\i}a${}^1$\thanks{~Email: garcia@cpt.univ-mrs.fr}$\;$,
        R. Trinchero${}^2$\thanks{~Email: trincher@cab.cnea.gov.ar} \\
\\
${}^1$ {\it Centre de Physique Th\'eorique - CNRS}                  \\
       {\it Campus de Luminy, Case 907, F-13288 Marseille, France}  \\
\\
${}^2$ {\it Instituto Balseiro and Centro At\'omico Bariloche}      \\
       {\it CP 8400 - Bariloche, R\'{\i}o Negro, Argentina}         \\
\\
}

\date{}

\begin{document}

\begin{titlepage}
\thispagestyle{empty}

\maketitle

\begin{abstract}
We define a quantum generalization of the algebra of functions over an 
associated vector bundle of a principal bundle. Here the role of a quantum 
principal bundle is played by a Hopf-Galois extension. Smash products of 
an algebra times a Hopf algebra $H$ are particular instances of these 
extensions, and in these cases we are able to define a differential 
calculus over their associated vector bundles without requiring the use 
of a (bicovariant) differential structure over $H$. Moreover, if $H$ is 
coquasitriangular, it coacts naturally on the associated bundle, and the 
differential structure is covariant.

We apply this construction to the case of the finite quotient of the 
$SL_q(2)$ function Hopf algebra at a root of unity ($q^3=1$) as the 
structure group, and a reduced $2$-dimensional quantum plane as both the 
``base manifold'' and fibre, getting an algebra which generalizes the 
notion of classical phase space for this quantum space. We also build 
explicitly a differential complex for this phase space algebra, and find 
that levels $0$ and $2$ support a (co)representation of the quantum 
symplectic group. On this phase space we define vector fields, and with 
the help of the $Sp_q$ structure we introduce a symplectic form relating 
$1$-forms to vector fields. This leads naturally to the introduction 
of Poisson brackets, a necessary step to do ``classical'' mechanics on 
a quantum space, the quantum plane.
\end{abstract}

\vspace{0.6 cm}

\noindent Keywords: non commutative geometry, fibre bundles,
                    quantum groups, symplectic structures. 

\vspace{0.5cm}

\noindent {\tt math-ph/9908007}\\
\noindent CPT-99/P.3883

\vspace*{0.4 cm}

\end{titlepage}
%%%%%%%%%%%%%%%%%%%%%%%%%%%%%%%%%%%%%%%%%%%%%%%%%%%%%%%%%%%%%%%%%%%%%%%%%%
%%%%%%%%%%%%%%%%%%%%%%%%%%%%%%%%%%%%%%%%%%%%%%%%%%%%%%%%%%%%%%%%%%%%%%%%%%

\section{Introduction}

The Gel'fand-Naimark theorem \cite{Gelfand} tells us that we can 
equivalently study a topological manifold through its algebra of 
(continuous) functions. Non commutative (NC) geometry builds upon this 
fact, considering arbitrary non commutative algebras as algebras of would 
be functions on hypothetical non commutative (or ``quantum'') spaces. In 
this context we have a standard description of vector fibre bundles as 
(finitely generated, projective) $A$-modules, where $A$ is the algebra 
encoding the ``base manifold''. Essentially, elements of this module 
should be thought as sections of the fibre bundle. In the commutative 
(classical) case, this correspondence encodes all the structure of the 
bundle, as asserted by the Serre-Swan theorem. 

However, any bundle is, in particular, a manifold (the ``total space'').
Therefore, in the quantum case we should expect its noncommutative version 
to be encoded by an algebra. Of course, in the same way that the ``total''
manifold is obviously not enough to define a classical bundle, in the 
quantum case we will need an algebra with additional structure. One of the 
physical motivations for trying to characterize a quantum vector bundle 
using an algebra is the following. One could try to generalize the standard 
way of doing classical mechanics to some quantum space. But this needs first 
the construction of a phase space (that means, the cotangent bundle) over 
such NC space, and then of a differential algebra on this phase space. 
Hence, having a module as a phase space is not enough to reach this aim, and 
we again see the need of an algebra to encode the structure of the total 
space for such a vector bundle. It should be noted here that an alternative 
---and more direct--- approach is the one taken in \cite{Albeverio}, where
the starting quantum space is considered as a phase space.

As principal fibre bundles are characterized, in NC geometry, by 
Hopf-Galois extensions (algebras), we first construct (section~2) a 
quantum generalization of a vector bundle as associated vector bundle to 
a principal bundle. We illustrate this construction with the example of a 
quantum plane. Although developed independently, this definition of quantum 
vector bundles turned out to be essentially the same as the one introduced 
in \cite{Brzezinski-Majid} and \cite{Budzynski-Kondracki}. In section~3 
we write down the definition of a differential complex over such bundles. 
The second part of this work (section~4) is devoted to the definition of 
a symplectic structure on a quantum bundle over a reduced quantum plane, 
namely the phase space for the $2D$ reduced quantum plane at $q^3=1$. We 
first define vector fields, their pairing with $1$-forms, and a symplectic
$2$-form. Finally, we define Poisson brackets and we mention some of its
properties.

%%%%%%%%%%%%%%%%%%%%%%%%%%%%%%%%%%%%%%%%%%%%%%%%%%%%%%%%%%%%%%%%%%%%%%%%%%
%%%%%%%%%%%%%%%%%%%%%%%%%%%%%%%%%%%%%%%%%%%%%%%%%%%%%%%%%%%%%%%%%%%%%%%%%%

\section{Quantum bundles}

Let $H$ be a Hopf algebra, and $B \subset P$ two algebras. One says that 
$P$ is a (right) extension of $B$ by $H$ if $P$ is a (right) $H$-comodule 
algebra such that
$$
B = P^{CoH} = \left\{ b\in P\;/\;\delta_R b = b\otimes \one_H\right\} \ .
$$
In particular, this means that we can always say that an $H$-comodule 
algebra $P$ is an $H$-extension of its coinvariant subalgebra $P^{CoH}$. 
Moreover, if the map
\begin{eqnarray*}
   \beta : P\otimes_B P &\longmapsto & P\otimes H \\
   \beta &=& (m_P \otimes \id)\circ (\id \otimes \delta _{R}) \\
   \beta \left( p\otimes_{B} p^{\prime }\right) 
      &\equiv & pp_0^\prime\otimes p_1^\prime \quad
      \text{with} \quad \delta_R p^\prime = p_0^\prime \otimes p_1^\prime 
                                   \in P\otimes H 
\end{eqnarray*}
is bijective, one says that $P$ is a (right) Hopf-Galois extension of 
$B$ by $H$. 

Classically, a Hopf-Galois extension is the dual object to a principal fibre 
bundle. In fact, in such a case the algebra $B$ is simply the algebra of 
functions on the base manifold, $P$ the algebra of functions on the total 
space, and $H$ the algebra of functions on the structure group. The 
condition involving the $\beta$ function encodes the fact that all the 
fibres are isomorphic to the structure group that acts freely on the bundle.

For $A$ a left $H$-module algebra, the {\em smash} product algebra $A\#H$ 
is defined to be the tensor product $A\otimes H$ as a vector space, with 
the modified multiplication
\begin{equation}
(a\#h)(b\#g)\equiv a(h_1\triangleright b)\,\#\,h_2 g \quad ,
   \quad \quad a,b\in A\,,\quad h,g\in H \ ,
\end{equation}
where $h_1 \otimes h_2 =\Delta h$ is the coproduct. Such a smash product 
is automatically a right $H$-comodule algebra, with
\begin{equation}
   \delta_R(a\#h) \equiv (a\#h_1 )\otimes h_2 \ .
\end{equation}
Moreover, this is a particular case of a (right) Hopf-Galois extension of 
$A$ by $H$, as it can be easily seen. In fact, it corresponds to a 
particular case of a globally trivial (quantum) principal bundle, as it 
has a (dual) analogue of a global section.
General trivial bundles are given by {\em cross} product algebras 
\cite{Majid}, in which smash product data is replaced by a weakened form 
through the introduction of a cocycle (be aware that some authors just 
call cross product algebras what we have here called smash ones). In the 
commutative ``limit'', the action of $H=C(G)$ on $A=C(M)$ reduces to the 
trivial action and we get back the standard tensor product algebra of 
functions on a globally trivial $M\times G$ manifold.

Suppose that we now want to build an associated vector bundle, given some 
principal bundle $E$. Classically, being $G$ the structure group of $E$, 
one first chooses a vector space $V$ with a left $G$-action. Then the 
associated vector bundle $F$ with fibre $V$ is defined to be the quotient 
of $E \times V$ by the equivalence relation 
$(e,gv) \sim (eg,v) \:, \ e\in E, v \in V, g\in G$.

Dually, we need a left $H$-comodule algebra $W$ playing the role of the 
algebra $C(V)$. Now, the associated vector bundle will be encoded by 
\begin{equation}
   Q \equiv P \stackunder{H}{\Box} W
\end{equation}
which corresponds to the algebra of functions on $F$. Here the cotensor 
product $P \stackunder{H}{\Box} W$ is defined to be
$$
   P \stackunder{H}{\Box} W \equiv \left\{ \;p_i\otimes w_i \in 
              P \otimes W\;\diagup \; 
              \delta_R \, p_i\otimes w_i=p_i\otimes \delta_L w_i \right\} \ .
$$
As it can be checked easily, this subspace of $P \otimes W$ is invariant 
under the product of the tensor product algebra. Therefore, 
$P \stackunder{H}{\Box} W$ is in fact an algebra, with
$$
   (p\otimes w)(p'\otimes w')=(pp'\otimes ww') 
   \qquad \text{whenever} 
   p\otimes w, p'\otimes w' \in P \stackunder{H}{\Box} W \ .
$$

If we take $P$ to be a smash product, $P= A\# H$, then we will have
$Q = P \stackunder{H}{\Box} W = A\# H \stackunder{H}{\Box} W$, a vector 
subspace of $A \otimes H \otimes W$. But we see that the left coaction 
$\delta_L$ on $W$ is a vector space isomorphism between 
$H \stackunder{H}{\Box} W$ and $W$, and hence $\id \otimes \delta_L$ is 
a vector space isomorphism between $A \otimes H \stackunder{H}{\Box} W$ 
and $A \otimes W$. Using this isomorphism we can now map the general 
product found in $Q$ to this isomorphic space, finding that $A \otimes W$ 
is an algebra ---that we call $A \Box W$--- with 
$$
   (a\Box v)(b\Box w) =
      a\,(v_{-1}\triangleright b)\Box \,v_{0}\,w \ .
$$
So, in this particular case we will simply take $Q=A \Box W$ as the 
algebra encoding the associated vector bundle. We also mention that, if 
$H$ is coquasitriangular with universal $r$-form $r$, there is a left 
coaction $\Delta_L$ of $H$ on $Q$ given by
$$
   \Delta_L \, (a\Box w) = a_{-1}\, w_{-1} \otimes (a_0\Box w_0) \ .
$$
Here, the left $H$-comodule structure on $A$ 
($\delta_L \, a = a_{-1} \otimes a_0$) should be related to the left 
$H$-action on $A$ by coquasitriangularity,
$$
   h \triangleright a \equiv r(a_{-1}\otimes h)\,a_0 \ .
$$

%%%%%%%%%%%%%%%%%%%%%%%%%%%%%%%%%%%%%%%%%%%%%%%%%%%%%%%%%%%%%%%%%%%%%%%%%%
%%%%%%%%%%%%%%%%%%%%%%%%%%%%%%%%%%%%%%%%%%%%%%%%%%%%%%%%%%%%%%%%%%%%%%%%%%

\subsection{Phase space for a quantum plane}

As an example of the previous construction, and to use it afterwards, we 
define here a bundle over a quantum plane. Let $q$ be a cubic root of unity, 
$q^3=1$, and call $\mathcal{M}$ the reduced quantum plane defined by
\begin{equation}
   xy=q\,yx \ , \qquad x^3=y^3=\one \ .
\label{product-on-M}
\end{equation}
In order to obtain a ``phase space'' over $\mathcal{M}$ we obviously need 
to take $A=\mathcal{M}$. By analogy with the classical case we also take 
$W=\mathcal{M}$. Therefore the natural candidate for $H$ is the quantum 
group ${\mathcal{F}}=Fun(SL_q(2,\CC)) \: / \: a^3=d^3=\one\;,\;b^3=c^3=0$ 
---and not its dual, see the discussion about classical counterparts above.
We refer to \cite{CoGaTr-l,CoGaTr-e} for a much more detailed analysis of 
such objects and further references on the subject, and in particular for
explicit formulas of the Hopf structure on $\mathcal{F}$ according to our
conventions. See also \cite{Hajac}, where tangent and cotangent bimodules 
associated to a Hopf-Galois extension of the $2d$ quantum plane by the
biparametric quantum group $Fun(GL_{q,p}(2))$ are built and analysed. 
Obviously, this bigger Hopf algebra can be reduced to $\mathcal{F}$ 
suitably choosing the parameter $p$ and taking the corresponding quotient 
when $q$ is a root of unit; therefore this construction provides a pair 
of tangent/cotangent bimodules for $\mathcal{M}$. Actually, the algebra
$Fun(SL_q(2,\CC))$ itself is a Hopf-Galois extension by $\mathcal{F}$ of 
its classical counterpart $Fun(SL(2,\CC))$; this was mentioned by 
\cite{Connes} and explicitly shown in \cite{Dabrowski}. However, we 
do not use this interesting property in the present paper.

Now we need both a left coaction and a left action of $\mathcal{F}$ on 
$\mathcal{M}$. The coaction we take is the standard one, given by
{\small
$$
\delta_L\left( 
  \begin{array}{c}
    x \\ y
  \end{array} \right) = \left(
  \begin{array}{cc}
    a & b \\
    c & d
  \end{array} \right) \dot{\otimes} \left(
  \begin{array}{c}
    x \\ y
  \end{array} \right) \ .
$$
}

\noindent
Of course, for the action we could take the trivial one, but that would 
reduce the smash product $\mathcal{M} \# \mathcal{F}$ to a tensor 
product\ldots the results would be uninteresting. However, using the fact 
that $\mathcal{F}$ is coquasitriangular (because $Fun(SL_q(2,\CC))$ is so 
\cite{Klimyk}) with universal $r$-form 
$r(T_{i}^{\;j}\otimes T_{k}^{\;l})\equiv q \,R_{ik}{}^{jl}$, 
we can map left comodules to left modules. Doing so with the comodule
($\mathcal{M}$, $\delta_L$) we get the following non-trivial left action 
of $\mathcal{F}$ on $\mathcal{M}$:
{\small
$$
\begin{tabular}{c|cc}
$\ \triangleright\ $ & $x$ & $y$ \\
\hline
$a$ & $\;q^{2}\,x\;$ & $q\,y$ \\
$b$ & $0$ & $(q^{2}-1)\,x$ \\
$c$ & $0$ & $0$ \\
$d$ & $q\,x$ & $q^{2}\,y$
\end{tabular}
$$
}

Just to simplify the formulas, we introduce a more compact notation 
for the generators of $\mathcal{M} \Box \mathcal{M}$, defining
\begin{eqnarray}
x \equiv x \Box \one & \qquad & p_x \equiv \one \Box x
    \label{notation-for-M_M} \\
y \equiv y \Box \one & \qquad & p_y \equiv \one \Box y \nonumber
\end{eqnarray}
Now the product (``commutation'') relations on 
$\mathcal{M} \Box \mathcal{M}$ may be written as
\begin{equation}
\begin{tabular}{p{5.5cm}p{0.7cm}p{5.4cm}}
\begin{eqnarray*}
  p_x \, x &=& q^2\,x p_x \\
  p_x \, y &=& q\,y p_x+(q^2-1)\,x p_y \\
  p_y \, x &=& q\,x p_y \\
  p_y \, y &=& q^2\,y p_y
\end{eqnarray*}
& &
\begin{eqnarray*}
  p_x \, p_y &=& q\,p_y\,p_x \\
  p_x^3 &=& \one \\
  p_y^3 &=& \one \\
\end{eqnarray*}
\end{tabular}
\label{product-on-M_M}
\end{equation}

\noindent
These relations are reminiscent of those of the Wess-Zumino (reduced) 
complex \cite{Wess-Zumino, CoGaTr-l}, but they are {\it not} the same. 
Of course, due to the notation (\ref{notation-for-M_M}), the relations 
(\ref{product-on-M}) are still valid in this bigger algebra. We also note 
that, disregarding the cubic relations, this algebra has the same 
Poincar\'e series as the commutative one (functions over a 
$2+2$-dimensional phase space).

%%%%%%%%%%%%%%%%%%%%%%%%%%%%%%%%%%%%%%%%%%%%%%%%%%%%%%%%%%%%%%%%%%%%%%%%%%
%%%%%%%%%%%%%%%%%%%%%%%%%%%%%%%%%%%%%%%%%%%%%%%%%%%%%%%%%%%%%%%%%%%%%%%%%%

\section{Differential calculus on a vector bundle}

Now we will show that it is possible to build a differential calculus 
over a quantum vector bundle of the type $Q=A\Box W$, making use only of 
covariant differential calculi $\Omega(A),\Omega(W)$ ---that is, no forms 
``on the quantum group $H$'' are ever needed, as would be the case for a 
more general $Q$.

As we said, we need
\begin{itemize}
\item[-] $\Omega(A)$ a left $H$-module differential calculus, 
  $\ h\triangleright d\,a = d(h\triangleright a)$.
\item[-] $\Omega(W)$ a left $H$-comodule differential calculus,
  $\ \delta _{L}(dw) = (id\otimes d)\delta _{L}w$.
\end{itemize}

\noindent 
Since $A\Box W$ coincides with $A\otimes W$ as a vector space, we 
also take as a vector space the following equality:
$$
   \Omega^1(Q)\stackunder{vs}{=} \Omega^1(A\otimes W)
              =\Omega^1(A) \otimes W \:\oplus\: A\otimes \Omega^1(W) \ .
$$
The differential operator $D: Q \mapsto \Omega^1(Q)$ should 
be taken as
$$
   D(a\Box w)=da\Box w+a\Box \,dw \ .
$$
Only the bimodule structure should be made non-trivial, to reflect the 
fact that $Q=A\Box W$ has a product which is different from the one in 
$A\otimes W$ (but similar to the product needed in $\Omega(A\# H)$, see
\cite{Pflaum}). The good choice is (the linear extension of)
\begin{eqnarray*}
(a\Box v)\,\left( db\Box w+c\Box \,dt\right) &\equiv
  & a\,(v_{-1}\triangleright db)\Box \,v_{0}\,w
    + a\,(v_{-1}\triangleright c)\Box \,v_{0}\,dt \\
\left( db\Box w+c\Box \,dt\right)\,(a\Box v) &\equiv
  & db\,(w_{-1}\triangleright a)\Box \,w_{0}\,v
    + c\,(t_{-1}\triangleright a)\Box \,dt_{0}\,v
\end{eqnarray*}

The generalization to a higher order differential calculus is pretty 
straightforward. As vector spaces one should take
\begin{eqnarray*}
  \Omega (A\Box W) &\equiv &
     \stackunder{n\geqslant 0}{\bigoplus }\Omega^n(A\Box W) \\
  && \Omega^n(A\Box W)\stackunder{vs}{\equiv}\;
     \stackrel{n}{\stackunder{k=0}{\bigoplus}}\,
     \left[ \Omega^{n-k}(A)\otimes \Omega^k(W)\right]
\end{eqnarray*}
The differential operator is graded, so we must set 
\begin{eqnarray}
D: \Omega^{n}(A\Box W) &\longrightarrow & \Omega ^{n+1}(A\Box W) \\
D(\alpha \Box \,\omega )&=&
    d\alpha \,\Box \,\omega + (-1)^k\,\alpha \Box\,d\omega \ ,
    \qquad \alpha \in \Omega^k(A)\,,\;\omega \in \Omega^{n-k}(W) \ . 
    \nonumber
\end{eqnarray}

\noindent
We also need a product for this differential calculus, extending the 
bimodule structure already written:
\begin{eqnarray}
(\alpha \Box \nu )(\beta \Box \omega) &=&
   (-1)^{(n-k)j}\,\alpha (\nu_{-1}\triangleright \beta)
       \Box\,\nu_0\omega \\
   & & \alpha \in \Omega^k(A)\,,\;\nu \in \Omega^{n-k}(W)\,,\;\
   \beta \in \Omega^j(A)\,,\;\omega \in \Omega^{m-j}(W) \ . \nonumber
\end{eqnarray}

%%%%%%%%%%%%%%%%%%%%%%%%%%%%%%%%%%%%%%%%%%%%%%%%%%%%%%%%%%%%%%%%%%%%%%%%%%
%%%%%%%%%%%%%%%%%%%%%%%%%%%%%%%%%%%%%%%%%%%%%%%%%%%%%%%%%%%%%%%%%%%%%%%%%%

\subsection{Differential algebra on $\mathcal{M}\Box \mathcal{M}$}

Let's now show explicit formulas for our example, the differential algebra
$\Omega(\mathcal{M} \Box \mathcal{M})$ on the phase space of the quantum 
plane $\mathcal{M}$. To this end we obviously need the differential 
algebra $\Omega(\mathcal{M})$ \cite{Wess-Zumino, CoGaTr-l}. It is given by
\begin{equation}
\begin{tabular}{p{5.5cm}p{0.7cm}p{5.4cm}}
\begin{eqnarray*}
  x\,dx &=& q^2 dx\,x \\
  x\,dy &=& q\,dy\,x+(q^2-1)\,dx\,y \\
  y\,dx &=& q\,dx\,y \\
  y\,dy &=& q^2 dy\,y
\end{eqnarray*}
& &
\begin{eqnarray*}
  dx^2 &=& 0 \\
  dy^2 &=& 0 \\
  dx\,dy &=& - q^2 dy\,dx \\
\end{eqnarray*}
\end{tabular}
\end{equation}

\noindent
Here we use again the shortened notation, calling $dz \Box \one = dz$, 
$\one\Box dz = dp_z$, with $z=x,y$. In this way, the product relations 
happen to be those found at level zero, already shown in 
(\ref{product-on-M_M}), plus the following level $1$ equalities,
\begin{equation}
\begin{tabular}{p{5.7cm}p{0.7cm}p{6.2cm}}
\begin{eqnarray*}
  x\,dx &=& q^2 dx\,x \\
  y\,dx &=& q \,dx\,y \\
  x\,dy &=& q \,dy\,x + (q^2-1)\,dx\,y \\
  y\,dy &=& q^2 dy\,y \\
  && \\
  dp_x\,x &=& q^2 x\,dp_x \\
  dp_x\,y &=& q \,y\,dp_x + (q^2-1)\,x\,dp_y \\
  dp_y\,x &=& q \,x\,dp_y \\
  dp_y\,y &=& q^2 y\,dp_y
\end{eqnarray*}
& &
\begin{eqnarray*}
  p_x\,dp_x &=& q^2 dp_x\,p_x \\
  p_y\,dp_x &=& q \,dp_x\,p_y \\
  p_x\,dp_y &=& q \,dp_y\,p_x + (q^2-1)\,dp_x\,p_y \\
  p_y\,dp_y &=& q^2 dp_y\,p_y \\
  && \\
  dx\,p_x &=& q \,p_x\,dx \\
  dx\,p_y &=& q^2 p_y\,dx \\
  dy\,p_x &=& q^2 p_x\,dy + (q-1)\,p_y\,dx \\
  dy\,p_y &=& q \,p_y\,dy
\end{eqnarray*}
\end{tabular}
\label{product-on-Omega_1}
\end{equation}

\noindent
and finally the level $2$ ones,
\begin{equation}
\begin{tabular}{p{4.4cm}p{0.7cm}p{6.6cm}}
\begin{eqnarray*}
  dx^2       &=& dy^2   = 0 \\
  dp_x^2     &=& dp_y^2 = 0 \\
  dx\,dy     &=& - q^2 dy\,dx \\
  dp_x\,dp_y &=& - q^2 dp_y\,dp_x
\end{eqnarray*}
& &
\begin{eqnarray*}
  dp_x\,dx &=& -q^2 dx\,dp_x \\
  dp_x\,dy &=& -q \,dy\,dp_x + (1-q^2)\,dx\,dp_y \\
  dp_y\,dx &=& -q \,dx\,dp_y \\
  dp_y\,dy &=& -q^2 dy\,dp_y
\end{eqnarray*}
\end{tabular}
\label{product-on-Omega_2}
\end{equation}

\noindent
Hence, any monomial can be reordered in a predetermined way, and 
$\Omega(\mathcal{M} \Box \mathcal{M})$ has the same Poincar\'e 
series as its classical counterpart.

Notice that the algebra spanned by $\{ dx,dy\}$ is the Manin dual of the 
algebra spanned by $\{ x,y\}$, and the same is true for $\{ dp_x,dp_y\}$ 
and $\{ p_x,p_y\}$. However, we should stress the fact that this fails 
to be true for $\{ dx,dy,dp_x,dp_y\}$ and $\{ x,y,p_x,p_y\}$.

%%%%%%%%%%%%%%%%%%%%%%%%%%%%%%%%%%%%%%%%%%%%%%%%%%%%%%%%%%%%%%%%%%%%%%%%%%
%%%%%%%%%%%%%%%%%%%%%%%%%%%%%%%%%%%%%%%%%%%%%%%%%%%%%%%%%%%%%%%%%%%%%%%%%%

\section{Symplectic structure on $\mathcal{M}$}

\subsection{Symplectic q-group on phase space}

In can be checked that levels $0$ and $2$ of the differential complex
$\Omega(Q)$ on the phase space $Q=\mathcal{M}\Box\mathcal{M}$ (relations
(\ref{product-on-M},\ref{product-on-M_M},\ref{product-on-Omega_2}))
are corepresentation spaces for the $Sp_q(2)$ quantum group (in 
addition to being also $SL_q(2)$ covariant!). In fact, using the 
following identifications,
\begin{equation}
\begin{tabular}{p{4.0cm}p{0.8cm}p{4.0cm}}
\begin{eqnarray*}
  x \ &\leftrightarrow &x_1 \\
  y \ &\leftrightarrow &x_2 \\
  p_x &\leftrightarrow &x_3 \\
- p_y &\leftrightarrow &x_4
\end{eqnarray*}
& &
\begin{eqnarray*}
  dp_x &\leftrightarrow &\xi_1 \\
  dp_y &\leftrightarrow &\xi_2 \\
    dx &\leftrightarrow &\xi_3 \\
  - dy &\leftrightarrow &\xi_4
\end{eqnarray*}
\end{tabular}
\end{equation}

\noindent
the product relations at levels $0$ and $2$ of the differential algebra 
$\Omega(Q)$ can be rewritten
\begin{eqnarray*}
  (\hat{R}_{Sp}-q)\,(x\otimes x) &=& 0 \\
  \left(\hat{R}_{Sp}^{\;2}-\hat{R}_{Sp}+\one\right) 
               (\xi \otimes \xi) &=& 0 \ .
\end{eqnarray*}
Here 
$\left(\hat{R}_{Sp}\right)_{ij}^{\;kl} = \left(R_{Sp}\right)_{ji}^{\;kl}\:$,
and $R_{Sp}$ is the $R$-matrix of the $Sp_q(2)$ quantum group to be found 
in \cite{FRT}. Remark that here $\xi_i \neq dx_i$\ldots this is why this 
symmetry is not found at level $1$.

%%%%%%%%%%%%%%%%%%%%%%%%%%%%%%%%%%%%%%%%%%%%%%%%%%%%%%%%%%%%%%%%%%%%%%%%%%
%%%%%%%%%%%%%%%%%%%%%%%%%%%%%%%%%%%%%%%%%%%%%%%%%%%%%%%%%%%%%%%%%%%%%%%%%%

\subsection{Vector fields}

In differential geometry, vector fields are defined as the $C(M)$-module 
of derivations $Der(C(M))$ on $C(M)$. However, this definition is not 
longer a good one for the quantum case, as the space $Der(A)$ would only 
be a module over the center of the corresponding algebra $A$. So this 
gives rise to different possibilities for defining vector fields. But 
having already a differential algebra $\Omega(A)$ for a quantum space, 
the natural thing to do is to build vector fields as objects dual to 
$1$-forms. In general, they happen to be {\em twisted} derivations.

In the same way as in the case of the $2d$ quantum plane 
\cite{Wess-Zumino}, here we define derivative operators on the phase 
space $Q=\mathcal{M}\Box \mathcal{M}$ as the ones such that
\begin{equation}
  d = dx\,\partial_x + dy\,\partial_y + 
      dp_x\,\partial_{p_x} + dp_y\,\partial_{p_y} \ .
\label{d_operator_with_derivatives}
\end{equation}
The operators $\partial$ exist and are well defined because any $1$-form
can be expanded in a unique way using $dx,dy,dp_x,dp_y$ as generators of 
$\Omega^1(Q)$ as a right $Q$-module. More generally, one can define in a 
similar way derivations for an associative algebra $Q$ such that the 
uniqueness condition of the right expansion applies (replacing 
$dx,dy,dp_x,dp_y$ by a set of generators $\{dz_i\}$ of $\Omega^1(Q)$ as a 
$Q$-bimodule). The derivatives $\partial_.$ happen to be {\em twisted} 
derivative operators, as they will generally not satisfy Leibniz's rule.
Operators $\partial_.$ naturally generate a $Q$-(bi)module, remembering 
that elements of $Q$ may also be thought as (multiplicative) operators 
$Q \mapsto Q$. This space ${\mathcal{D}}^1$ is the bimodule of first 
order differential operators. We also introduce a bilinear mapping 
$\langle \:,\,\rangle : \Omega^1 \otimes {\mathcal{D}}^1 \mapsto Q$,
pairing one-forms and vector fields as follows:
$$
\langle df\,g\,,\,X\rangle \equiv X(f)\,g\;,\qquad
   f,\,g\in Q\;,\qquad X\in {\mathcal{D}}^1 \ .
$$
One can show that this pairing is well defined, given the choice in 
(\ref{d_operator_with_derivatives}).

%%%%%%%%%%%%%%%%%%%%%%%%%%%%%%%%%%%%%%%%%%%%%%%%%%%%%%%%%%%%%%%%%%%%%%%%%%

\subsection{Symplectic $2$-form}

Classically, a symplectic form is a closed non-degenerate $2$-form. The 
first condition may be automatically fulfilled in the $NC$ context by 
taking $\omega = w_{ij}\,\xi_i\,\xi_j$,
with $w_{ij}\in \CC$. Moreover, higher order forms are classically defined 
to be wedge (antisymmetric) products of $1$-forms, and a pairing of such 
forms with vector fields is defined componentwise. On the contrary, here
we are using forms which do not correspond to the antisymmetrization 
of a product of $1$-forms. Therefore this fact should be taken care of in 
the pairing with vector fields. So, we define the evaluation of a symplectic 
form $\omega$ on a pair of two vector fields using 
$A_{Sp}=(\hat{R}_{Sp}^{\;2}-q^2)/(q-q^{-1})$, the $Sp_q(2)$ antisymmetrizer 
(classically, it enters in the wedge product):
$$
\omega (X,Y)\equiv 
  w_{ij}\,A_{ij,kl}\,\langle \xi_k\,,\,X\rangle \,
                     \langle \xi_l\,,\,Y\rangle \ .
$$
In this way equivalent expressions of $\omega$ obtained by reordering 
$\xi_i,\xi_j$ using (\ref{product-on-Omega_2}) result in the same value 
of the above pairing with $X$ and $Y$, since the symmetrizer
$S_{Sp}=-(q^2/2)\left(\hat{R}_{Sp}^{\;2}-\hat{R}_{Sp}+\one\right)$ is 
involved in such a reordering. Now, any closed $\omega$ provides a map 
between vector fields and $1$-forms, through the relation
\begin{eqnarray*}
   \omega (X_{df}\,,\,\cdot \,) &=&\langle df\,,\,\cdot \,\rangle \ .
\end{eqnarray*}

\noindent
The non-degeneracy of the symplectic form is implemented by requiring 
this mapping to be one-to-one \cite{Dubois-Violette}.

Notice that, for an arbitrary associative algebra $Q$, symplectic forms 
associated to the differential calculus $\Omega_{Der}(Q)$ (see 
\cite{Dubois-Violette}) lead to Poisson brackets that are distinct from 
those described here.

For the case $Q=\mathcal{M}\Box \mathcal{M}$ we can simply chose the 
non-degenerate $2$-form 
$$
   \omega =q\left(dx\,dp_x+dy\,dp_y\right) \ ,
$$
and we get
\begin{eqnarray*}
  X_{dx} = q^2 \,\partial_{p_x} & \qquad & X_{dp_x} = -q \,\partial_x \\
  X_{dy} = q^2 \,\partial_{p_y} & \qquad & X_{dp_y} = -q \,\partial_y \ .
\end{eqnarray*}

%%%%%%%%%%%%%%%%%%%%%%%%%%%%%%%%%%%%%%%%%%%%%%%%%%%%%%%%%%%%%%%%%%%%%%%%%%
%%%%%%%%%%%%%%%%%%%%%%%%%%%%%%%%%%%%%%%%%%%%%%%%%%%%%%%%%%%%%%%%%%%%%%%%%%

\subsection{Poisson brackets}

Having the $2$-form $\omega$, and a way to evaluate it on a pair of 
vector fields, we can now introduce Poisson brackets on $Q=A\Box W$ by
\begin{eqnarray*}
\left\{ f\,,\,g\right\} &\equiv &\omega (X_{df}\,,\,X_{dg})
   \qquad \qquad \qquad 
   f,\,g\in \mathcal{M}\Box \mathcal{M} \\
               &=& \langle df \,,\, X_{dg} \rangle =X_{dg}(f) \ .
\end{eqnarray*}

\noindent
The last expression makes it evident that these Poisson brackets are 
{\em twisted} derivations, considered as operators on their first
variable, since $X_{dg}$ itself is so.

In the case $Q=\mathcal{M} \Box \mathcal{M}$, the brackets amongst 
generators are easy to get, and one finds that the only non-zero ones are
\begin{eqnarray*}
   \left\{ x\,,\,p_{x}\right\} =-q & \qquad 
           & \left\{ p_{x}\,,\,x\right\} =q^2\\
   \left\{ y\,,\,p_{y}\right\} =-q & \qquad 
           & \left\{ p_{y}\,,\,y\right\} =q^2
\end{eqnarray*}
It can also be checked that the brackets 
$\{ \one\,,\, f(x,y,p_x,p_y)\}$, $\{ f(x,y,p_x,p_y)\,,\, \one\}$,
and $\{ f(x,y)\,,\, g(x,y)\}$ are all vanishing.

%%%%%%%%%%%%%%%%%%%%%%%%%%%%%%%%%%%%%%%%%%%%%%%%%%%%%%%%%%%%%%%%%%%%%%%%%%
%%%%%%%%%%%%%%%%%%%%%%%%%%%%%%%%%%%%%%%%%%%%%%%%%%%%%%%%%%%%%%%%%%%%%%%%%%

\subsection{Final comments}

The Poisson brackets introduced above may be used to define an analogue 
of canonical equations of motion, after choosing a Hamiltonian function 
$h \in {\mathcal{M}}\Box{\mathcal{M}}$, by $\partial_t f = \{f,h\}$.
However, a real structure should be first defined on the phase space and 
its differential algebra, in order to select a {\em real} symplectic form
$\omega$. Moreover, we must remark that the ``time evolution'' determined 
by these equations is {\em twisted} ($\partial_t$ is not a derivation).

%%%%%%%%%%%%%%%%%%%%%%%%%%%%%%%%%%%%%%%%%%%%%%%%%%%%%%%%%%%%%%%%%%%%%%%%%%
%%%%%%%%%%%%%%%%%%%%%%%%%%%%%%%%%%%%%%%%%%%%%%%%%%%%%%%%%%%%%%%%%%%%%%%%%%
\bigskip

\subsection*{Acknowledgements} 

A.G. gratefully acknowledges the C.N.R.S. for financial support.

\bigskip

%%%%%%%%%%%%%%%%%%%%%%%%%%%%%%%%%%%%%%%%%%%%%%%%%%%%%%%%%%%%%%%%%%%%%%%%%%
%%%%%%%%%%%%%%%%%%%%%%%%%%%%%%%%%%%%%%%%%%%%%%%%%%%%%%%%%%%%%%%%%%%%%%%%%%

\end{document}